\documentclass[rnote]{aa}  
\usepackage{graphicx}
\usepackage{txfonts}

\usepackage{epsf}
\usepackage{natbib}
\usepackage{astro_bib_macro}
\usepackage{amssymb}
\bibpunct{(}{)}{;}{a}{}{,}

\newcommand{\kms}{km\,s$^{-1}$}
\newcommand{\fuse}{\emph{FUSE}}
\newcommand{\visir}{\emph{VISIR}}

\newcommand{\vrad}{$v_{\rm rad}$} 
 
\newcommand{\flux}{ergs\,s$^{-1}$\,cm$^{-2}$}

\newcommand{\Av}{$A_{\rm v}$}

\begin{document}
\title{Searching for molecular hydrogen mid-infrared emission in the
  circumstellar environments of Herbig Be stars}

   \author{C. Martin-Za{\"\i}di\inst{1}        
          \and
           E.F. van Dishoeck\inst{2,3}
           \and 
           J.-C. Augereau\inst{1}
          \and
           P.-O. Lagage\inst{4}
	   \and
           E. Pantin\inst{4} 
        }

   \offprints{C. Martin-Za{\"\i}di}

   \institute{Laboratoire d'Astrophysique de Grenoble, CNRS,
     Universit\'e Joseph-Fourier, UMR5571, Grenoble, France \\
     \email{claire.martin-zaidi@obs.ujf-grenoble.fr}
     \and
     Leiden Observatory, Leiden University, P.O. Box 9513, 2300 RA
     Leiden, the Netherlands
     \and 
     Max Planck Institut f\"ur Extraterrestrische Physik,
     Giessenbachstrasse 1, 85748 Garching, Germany
       \and 
       Laboratoire AIM, CEA/DSM - CNRS - Universit\'e Paris Diderot,
       DAPNIA/Service d'Astrophysique, Bat. 709, CEA/Saclay, 91191
       Gif-sur-Yvette Cedex, France} 

   \date{Received ... / Accepted ...}

  \abstract
  {Molecular hydrogen (H$_2$) is the most abundant molecule in the
    circumstellar (CS) environments of young stars, and is a key
    element in giant planet formation. The measurement of the H$_2$
    content provides the most direct probe of the total amount of CS
    gas, especially in the inner warm planet-forming regions of the
    disks. }
  {Most Herbig Be stars (HBes) are distant from the Sun and their
    nature and evolution are still debated. We therefore conducted
    mid-infrared observations of H$_2$ as a tracer of warm gas around
    HBes known to have gas-rich CS environments.}
  {We report a search for the H$_2$ S(1) emission line at 17.0348
    $\mu$m in the CS environments of 5 HBes with the high resolution
    spectroscopic mode of \visir\ ({\it ESO VLT Imager and
      Spectrometer for the mid-InfraRed}). }
  {No source shows evidence for H$_2$ emission at 17.0348
    $\mu$m. Stringent 3$\sigma$ upper limits on the integrated line
    fluxes are derived. Depending on the adopted temperature, limits
    on column densities and masses of warm gas are also estimated.
    These non-detections constrain the amount of warm ($>150~K$) gas
    in the immediate CS environments of our target stars to be less
    than $\sim 1-10 M_{\rm Jup}$.}
   {}

   \keywords{stars: circumstellar matter -- stars: formation -- stars:
     pre-main sequence -- ISM: molecules }

   \titlerunning{Searching for H$_2$ emission around Herbig Be stars}
   \authorrunning{C. Martin-Za\"{\i}di et al.}

   \maketitle
%

\section{Introduction}

Molecular hydrogen (H$_2$) is the most abundant molecule in the
environment of young stars. It remains optically thin up to high
column densities ($\sim$10$^{23}$ cm$^{−2}$), and does not freeze onto
dust grain surfaces. It furthermore self-shields efficiently against
photodissociation by far-ultraviolet (FUV) photons. Therefore, H$_2$
is the only molecule that can directly constrain the mass reservoir of
molecular gas in the circumstellar (CS) environment of pre-main
sequence stars. On the other hand, H$_2$ is one of the most
challenging molecules to detect. Electronic transitions occur in the
UV to which the Earth's atmosphere is opaque, and rotational and
rovibrational transitions at infrared (IR) wavelengths are faint
because of their quadrupolar origin. FUV spectroscopic observations of
H$_2$ absorption lines have provided evidence for the presence of cold
and warm excited H$_2$ around numerous Herbig Ae/Be stars (HAeBes) but
have not allowed us to determine the spatial distribution of the
observed gas \citep[e.g.][]{klr08a}. Detections of H$_2$ line emission
from disks around young stars with {\it ISO} \citep{Thi01} were
contradicted by ground-based observations \citep{Richter02, Sako05},
which showed that the observed emission could be due to the
surrounding cloud material. \cite{Carmona08} modeled the mid-IR H$_2$
lines originating in a gas-rich disk, seen face-on, surrounding a
Herbig Ae (HAe) star at 140 pc from the Sun. By assuming that the gas
and dust were well-mixed in the disk, a gas-to-dust ratio of about
100, and that $T_{gas}=T_{dust}$, those authors demonstrated that
mid-IR H$_2$ lines could not be detected with the existing
instruments. Indeed, they were unable to detect any H$_2$ mid-IR
emission line in their sample of 6 HAe stars. At the present time,
observations of H$_2$ mid-IR emission lines have been reported in 8 of
76 T\,Tauri stars observed with {\it Spitzer}
\citep{Lahuis07}. Although mid-IR windows are strongly affected by sky
and instrument background emission, the advent of high spectral and
spatial resolution spectrographs allows us to study the H$_2$ emission
from the ground, as demonstrated in the cases of two HAeBes,
\object{HD~97048} \citep{klr07a} and AB Aur \citep{Bitner07}, for
which particular conditions are required in their CS disks. The
analyses of these data usually assumes that the H$_2$ excitation is in
local thermodynamic equilibrium (LTE) and can thus be characterized by
a single excitation temperature, which should be close to the gas
temperature because of the low critical densities. The AB Aur
observations infer an H$_2$ gas temperature that is significantly
higher than the dust temperature.
\begin{table*}
\begin{center}
  \caption{Astrophysical parameters of the sample stars (columns 2 to
    6).  \vrad\ is the radial velocity of the star in the heliocentric
    rest frame. Columns 7 to 12 summarize the observations. The
    airmass and seeing intervals are given from the beginning to the
    end of the observations. \vspace{-0.4cm}}
\begin{tabular}{lcccccccccccccccccc}
  \hline
  \hline
  Star      & Sp.    & $T_{eff}$     & \Av         &  \vrad        & $d$            &  $t_{exp}$ & Airmass & Optical      & Standard & Airmass & Optical\\
            & Type     & (K)          & (mag)        & (\kms)       & (pc)            &  (s)      &         & Seeing   &  Star     &        & Seeing \\
            &     &           &         &       &             &       &         & ('')  &       &        &  ('') \\
  \hline
HD~98922  & B9       & 10470$^{(1)}$ & 0.34$^{(1)}$ & -15$^{(2)}$   & $>$540$^{(1)}$      & 2700     & 1.14-1.15  & 0.76-1.41  & HD~25025  & 1.024-1.025 & 0.69-0.71 \\
HD~250550 & B7       & 12800$^{(3)}$ & 0.57$^{(3)}$ & +31$^{(4)}$   & 606$\pm$367$^{(5)}$ & 3600     & 1.32-1.50  & 0.83-1.50  & HD~93813  & 1.018-1.023 & 0.91-1.05  \\
HD~259431 & B5       & 15900$^{(3)}$ & 0.88$^{(3)}$ & +43$^{(4)}$   & 290$\pm$84$^{(5)}$  & 2250     & 1.23-1.32  & 0.75-1.28  & HD~39425  & 1.12-1.14   & 1.51-1.62 \\
HD~76534  & B2       & 20000$^{(6)}$ & 0.80$^{(7)}$ & +17$^{(4)}$   & $>$160$^{(1)}$      &  3600     & 1.05-1.14  & 0.85-1.25  & HD~93813  & 1.014-1.017 & 0.92-0.96  \\
HD~45677  & B2/B1    & 21400$^{(1)}$ & 0.87$^{(1)}$ & +21.6$^{(8)}$ & 500$^{(9)}$         &  3450     & 1.02-1.08  & 0.69-1.40  & HD~25025  & 1.024-1.025 & 0.69-0.71   \\
  \hline
\end{tabular}
\begin{list}{}{} 
  References: $^{(1)}$ \cite{VdA98b}; $^{(2)}$ \cite{Acke05}; $^{(3)}$
  \cite{JCB03b}; $^{(4)}$ \cite{FINK_JAN84};
  $^{(5)}$\cite{Brittain07}; $^{(6)}$ \cite{klr04b}; $^{(7)}$
  \cite{valenti00}; $^{(8)}$ \cite{Evans67}; $^{(9)}$ \cite{Zorec98}
  as quoted in \cite{Cidale01}.  \vspace{-0.8cm} \\
\end{list}
\label{param}
\end{center}
\end{table*}

In contrast to HAes, the more distant Herbig Be stars (HBes) earlier
than B9 type have been poorly studied. As a consequence, the nature
and evolution of the CS material surrounding HBes is still a subject
of controversy. The Spectral Energy Distributions (SEDs) are very
different from one HBe to another \citep[e.g.][]{HILL92}. Some have
SEDs that are comparable to those observed for HAes, and have been
classified as group I or group II stars haboring disks according to
the classification of \citet[][]{Meeus01}. However, numerous HBes SEDs
with little IR excess have been modeled successfully by free-free
emission originating in a gaseous CS envelope, and compared with
rapidly rotating classical Be stars. In addition, the mid-IR emission
observed in HBes is generally not confined to optically thick disks
but originates in more complex environments such as remnant envelopes
or halos \citep{LEINERT01, POLOMSKI02, Vinkovic06}. This implies that
there are structural differences between HAes and the more luminous
HBes \citep{NATTA00, LEINERT01}. These results are fully consistent
with the more rapid evolution of the more massive HAeBes, which should
still be partially embedded in their natal cloud. On the other hand,
near-IR interferometric data of HBes have been interpreted using
flared disks models \citep[e.g.][]{Kraus07}.

We selected five HBe stars whose CS environments are rich in gas as
observed by the \fuse\ satellite \citep{klr08a}.  The selected stars
have IR excesses that are assumed to represent the presence of CS
dusty disks \citep[e.g.][]{HILL92, POLOMSKI02}, but the disk scenario
has not yet been clearly established. The aim of our study is to
detect the H$_2$ pure rotational emission line at 17.0348 $\mu$m in
the vicinity of these stars with the high resolution mode of the
VLT/\visir\ spectrograph \citep{Lagage04} and constrain the warm CS
gaseous component.

\section{Selection of the target  stars}
\label{targets}

The selection of the five target stars was motivated by the previous
observations of H$_2$ electronic lines in the FUV spectral range,
which showed that the CS environments of HBes were rich in cold and
warm H$_2$ \citep{klr08a}. The detection of H$_2$ absorption lines
provided evidence for the presence of cold ($\sim$100~K) and warm/hot
CS gas (up to 1500~K) around these stars. We first briefly describe
the CS environments of each of the five sample stars. The main
astrophysical parameters of the stars are summarized in
Table~\ref{param}.

\begin{table*}
\begin{center}
  \caption{3$\sigma$ upper confidence limits on the integrated fluxes
    and intensities of the S(1) line, and upper limits on the total
    column densities of H$_2$ and masses of H$_2$ as a function of the
    adopted temperature. $\lambda _{obs}$ is the expected position of
    the line in the observed spectra.}
\vspace{-0.3cm}
\begin{tabular}{lcccccccccccccccccc}
  \hline
  \hline   
  Star   & $\lambda _{obs}$ & Integrated flux      & Intensity                            & \multicolumn{3}{c}{$N({\rm H}_2)$ upper limits}                     & &  \multicolumn{3}{c}{H$_2$ mass upper limits$^{\rm (a)}$}     \\
  HD       &  ($\mu$m)       & (\flux)             & (ergs\,cm$^{-2}$\,s$^{-1}$\,sr$^{-1}$) &  \multicolumn{3}{c}{(cm$^{-2}$)}                  & &   \multicolumn{3}{c}{ ($M_{\rm Jup}$$\sim$10$^{-3}$~$M_{\odot}$)} \\
  \cline{5-7}  \cline{9-11} 
  &         &                         &                        &    150K  & 300K & 1000K                             & &     150 K &    300 K  &  1000 K \\

  \hline    
98922  & 17.0328 & $<$1.2$\times$10$^{-14}$ & $<$1.2$\times$10$^{-3}$ & 1.1$\times$10$^{23}$ & 6.9$\times$10$^{21}$ & 1.6$\times$10$^{21}$ & & 3.3               & 2.0$\times$10$^{-1}$               & 5.9$\times$10$^{-2}$  \\
250550 & 17.0373 & $<$1.6$\times$10$^{-14}$ & $<$3.8$\times$10$^{-3}$ & 1.5$\times$10$^{23}$ & 9.3$\times$10$^{21}$ & 2.2$\times$10$^{21}$ & & 5.5$_{-4.7}^{+8.8}$ & 3.4$_{-2.9}^{+5.4}$$\times$10$^{-1}$ & 1.0$_{-0.8}^{+1.6}$$\times$10$^{-1}$  \\
259431 & 17.0377 & $<$1.8$\times$10$^{-14}$ & $<$4.2$\times$10$^{-3}$ & 1.6$\times$10$^{23}$ & 1.0$\times$10$^{22}$ & 2.4$\times$10$^{21}$ & & 1.4$_{-0.7}^{+0.9}$ & 8.6$_{-4.3}^{+5.7}$$\times$10$^{-2}$ & 2.5$_{-1.3}^{+1.7}$$\times$10$^{-2}$ \\
45677  & 17.0365 & $<$1.6$\times$10$^{-14}$ & $<$3.7$\times$10$^{-3}$ & 1.4$\times$10$^{23}$ & 9.1$\times$10$^{21}$ & 2.1$\times$10$^{21}$ & & 3.7               & 2.3$\times$10$^{-1}$                & 6.7$\times$10$^{-2}$ \\
  \hline
\end{tabular}
\begin{list}{}{} 
  (a) Masses of H$_2$ are calculated assuming the distances quoted in
  Table~\ref{param} (see text). For HD~98922, the lower limit on the
  distance is used and the mass scales with the distance as
  $d^2$. \vspace{-0.8cm}
\end{list}
\label{tab_flux}
\end{center}
\end{table*}

{\bf {\object HD~98922}:} This star is classified as a self-shadowed
disk on the basis its SED \citep{van_Boekel03}. Its optical spectrum
displays broad, but relatively weak [O~I] emission profiles, which
cannot originate in the dusty disk surface but possibly in a rotating
gaseous disk inside the dust-sublimation radius \citep{Acke05}. Our
preliminary analysis of its \fuse\ spectrum indicates that the CS
environment of this star is rich in H$_2$. The excitation of H$_2$
appears to be similar to the HBes observed by \cite{klr08a}. The
modeling of the excitation conditions with the Meudon PDR Code
\citep{LePetit06} allows us to conclude that the observed gas is
probably located in the remnant cloud where the star formed. We
emphasize that no H$_2$ mid-IR line was detected by the Spitzer Space
Telescope towards this star \citep{Lahuis07}.

{\bf {\object HD~250550} and {\object HD~259431}:} The SEDs of these
two stars show signs of strong near-IR excesses that were modeled with
flat optically thick accretion disks by
\cite{HILL92}. Spectro-polarimetric observations provided indications
of flattened structures surrounding the stars
\citep{VINK02b}. However, spectroscopic observations provided no
evidence for Keplerian rotation with single-peaked emission lines
detected for both \ion{H}{I} Br$\gamma$ and CO \citep{Brittain07}. The
excitation of H$_2$ in the CS environments of these stars, revealed by
the \fuse\ satellite, favored an interpretation in terms of CS
envelopes which were the remnants of the parent molecular clouds in
which the stars were formed \citep{JCB03b, klr08a}.

{\bf {\object HD~76534}:} \cite{HILL92} analyzed the SED of HD~76534
and stressed that the low near-IR excess was probably due to free-free
emission in an ionized envelope rather than CS dust. By analyzing FUV
absorption lines of H$_2$ and deriving the excitation conditions of
the gas around the star, \cite{klr04b} concluded that the CS
environment of HD~76534, which is a B2 star, is probably too hostile
for the presence of a CS disk.

{\bf {\object HD~45677}:} The measured infrared excess was attributed
to a dust shell surrounding HD~45677 by \cite{Swings71}. The presence
of a disk was inferred by polarization and UV spectroscopic
measurements of accreting gas \citep{Schulte92, Grady93b}; these
studies also agreed with the presence of an actively accreting CS disk
around the star. The presence of such a disk is confirmed by the
analysis of emission lines from ionized metals by
\cite{Muratorio06}. Our preliminary analysis of the \fuse\ spectrum of
this star shows the presence of two or more gaseous components along
the line of sight. This may be consistent with the presence of a disk
associated with a CS envelope.

\section{Observations and data reduction}
\label{reduction}

Searches for mid-IR H$_2$ lines can be undertaken with the ESO/\visir\
instrument due to its high spectral and spatial resolution. Its high
spectral resolution (10\,000 $< R <$ 30\,000) allows us to disentangle
the H$_2$ lines from the absorption lines due to the Earth's
atmosphere.  \visir\ offers access to the most intense pure rotational
lines of H$_2$: the S(1) line at 17.0348 $\mu$m, S(2) at 12.2786
$\mu$m, S(3) at 9.6649 $\mu$m, and S(4) at 8.0250 $\mu$m. The S(0)
transition close to 28 $\mu$m is not observable from the ground due to
the Earth's atmospheric absorption, and the S(3) line, which is
located amidst a forest of telluric ozone features, is only observable
for particularly favorable Doppler shifts.

The five sample stars were observed with the high spectral resolution
long-slit mode of \visir\ on January 15 and 16, 2006. The central
wavelength of the observations was set to be 17.035 $\mu$m. We used
the 0.75'' slit, providing a spectral resolution of about 14\,000. The
details of the observation conditions are summarized in
Table~\ref{param}. According to its {\it IRAS} fluxes, HD~76534 should
be observable in the spectral range of \visir. However, during the
observation, the mid-IR flux of this star appeared to be very low and
the resulting spectrum could not be used to complete an accurate
analysis.

For all other observations, the standard ``chopping and nodding''
technique was used to suppress the large sky and telescope background
dominating at mid-IR wavelengths \cite[for details of the observation
technique see][]{klr07a}. The chopping was performed with a 8''
amplitude, which enabled us to detect the gas of any elongated
structure around our targets. To correct the spectrum for the Earth
atmospheric absorption and to obtain the absolute flux calibration, we
observed standard stars
\footnote{http://www.eso.org/sci/facilities/paranal/instruments/visir/tools/}
immediately before and after observing our sources. Since the HD~45677
and HD~98922 spectra have far higher signal-to-noise (S/N) ratios than
that of the standard stars, we divided the spectra of HD~98922,
HD~250550, and HD~259431 by that of HD~45677, and divided the spectrum
of HD~45677 by that of HD~98922, to correct for the telluric
absorption. We checked that the Doppler shifts were sufficiently
different from one star to the other, to ensure that the analysis
would not be significantly affected. We used the observed and modeled
spectra of the standard stars \citep{Cohen99} to derive the absolute
flux calibration. The wavelength calibration was completed by fitting
the observed sky background features with a model of Paranal's
atmospheric emission. The spectra were not corrected for dust
extinction, which is negligible in the mid-IR for \Av$<40$ mag
\citep{Fluks94}.

\section{Data analysis}
\label{analysis}

\begin{figure*}[!htbp] 
 \centering
  \includegraphics[width=14cm]{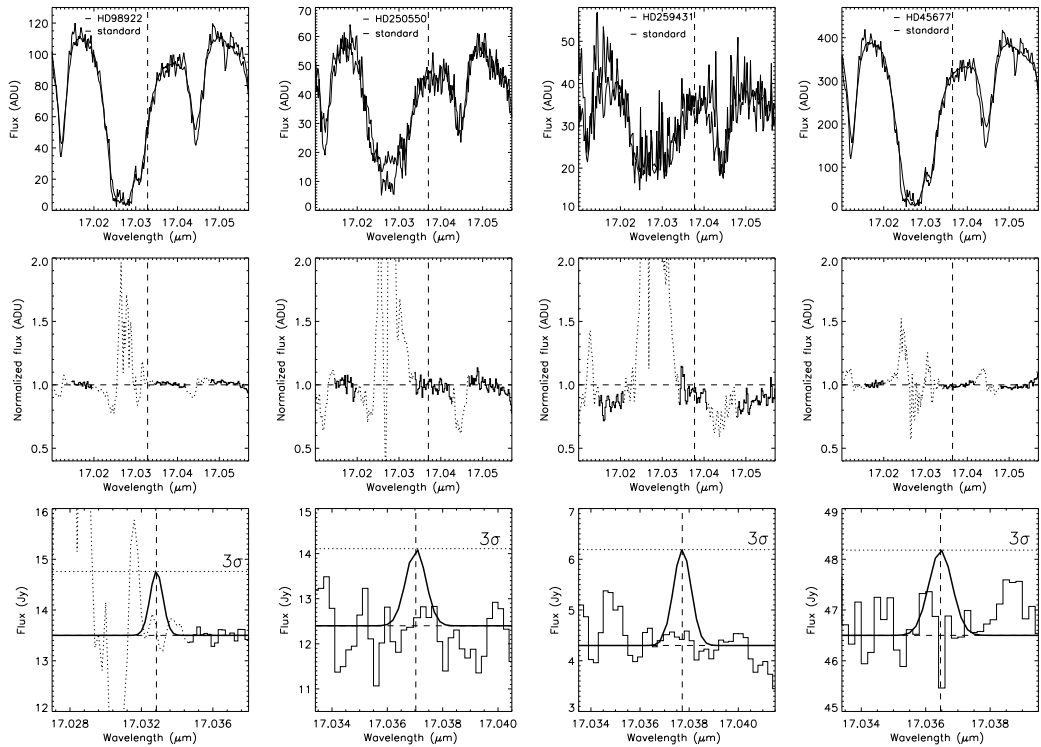}
  \caption{Spectra obtained for the H$_2$ S(1) line at 17.0348
    $\mu$m. {\it Top panel:} continuum spectra of the standard star
    and of the target before telluric correction. {\it Central panel:}
    full corrected spectra: dotted lines show spectral regions
    strongly affected by telluric features. {\it Bottom panel:} zoom
    of the region where the S(1) line should be observed (dashed
    vertical lines). A Gaussian of width FWHM = 21 \kms\ and
    integrated line flux equal to the 3$\sigma$ line-flux upper limits
    is overplotted. The spectra were corrected neither for the radial
    velocity of the targets nor the Earth's rotation velocity. }
  \label{spectres}
\end{figure*}

For the four high S/N spectra of stars in our sample, we display in
Fig.~\ref{spectres}, the spectra of the standard stars overplotted on
the target spectra (top panel), the entire atmosphere-corrected
spectra before the flux calibration (central panel), and the region of
the flux-calibrated atmosphere-corrected spectra where the S(1) H$_2$
line should appear if present (bottom panel). None of the observed
sources showed evidence for H$_2$ emission close to 17.035$\mu$m.

In the flux-calibrated spectra, we used two different methods to
determine the standard deviation ($\sigma$) of the continuum
flux. First, we calculated the standard deviation for wavelength
ranges relatively unaffected by telluric absorption, and close to the
wavelength of interest. Second, we estimated $\sigma$ in regions of
the spectra with comparable atmospheric transmission to the line of
interest. The first method provided more conservative values of
$\sigma$ for all but one star, namely HD~98922; for this star, the
line of interest should be detected at wavelengths corresponding to
the strong telluric absorption line, and the atmospheric transmission
played an important role. For our analysis, we kept the more
conservative values, and used the $\sigma$ values obtained with the
first method for HD~250550, HD~259431, and HD~45677, and the value
given by the second method for HD~98922.  The 3$\sigma$ upper limits
on the integrated line fluxes were calculated by integrating over a
Gaussian of full width at half maximum (FWHM) equal to a spectral
resolution element ($\Delta v \sim 21$ \kms), and an amplitude of
about 3$\times \sigma \times$ (flux), centered on the expected
wavelength for the S(1) line (Fig.~\ref{spectres}). The upper limits
on integrated fluxes and intensities of the S(1) line derived for each
star are tabulated in Table~\ref{tab_flux}. These limits were typically
a factor of 2 lower than the detection towards HD~97048
\citep{klr07a}. The intensities were calculated by dividing the
integrated fluxes by the solid angle of the slit of about 0.75'' width
with a spatial resolution element of about 0.427''.

From the upper limits on integrated intensities and by assuming that
the emitting H$_2$ is optically thin at LTE, we estimated the total
column densities of H$_2$ \citep[for details on the method,
see][]{Van_Dishoeck_92} as a function of adopted temperatures (150,
300, 1000~K). Even though the \fuse\ data indicated the presence of a
gas component with $T \sim$100~K for our target stars, we measured
150~K to be the lowest temperature of the gas, which can emit in the
H$_2$ S(1) line. For temperatures below 150~K, the S(0) line at 28
$\mu$m (not observable from the ground) provides more reliable
constraints on the optically thin gas properties than the S(1) line.

Under the same assumptions and assuming a homogeneous medium, the mass
of warm H$_2$ is given by:
$$
M_{warm~gas}=f\times 1.76\times 10^{-20}\frac{F_{ul}~ d^2}{(hc / 4 \pi
  \lambda)~ A_{ul}~ x_u(T)}~~~~M_{\odot} ,
$$
{\noindent}where $F_{ul}$ is the line flux, $d$ is the distance in pc
to the star, $\lambda$ is the wavelength of the transition $u-l$,
$A_{ul}$ is the spontaneous transition probability, $x_u(T)$ is the
fractional population of the level $u$ at the temperature $T$ in LTE,
and $f$ is the conversion factor required for deriving the total gas
mass from the H$_2$-ortho or H$_2$-para mass, assuming that ortho/para
ratio is controlled by the gas temperature \citep[from Eq.(1)
in][]{Takahashi01b}. Our results are presented in
Table~\ref{tab_flux}. The upper limits on the masses are calculated by
assuming the distances quoted in Table~\ref{param}. For HD~98922, we
used the lower limit on the distance and its mass scales with the
distance as $d^2$.

\section{Discussion}
\label{concl}

We observed five HBe stars with the high resolution spectroscopic mode
of \visir\ to search for H$_2$ pure rotational S(1) emission at
17.0348 $\mu$m. As probed by previous \fuse\ observations, the CS
environments of these stars provide evidence of large reservoirs
($N({\rm H}_2) \sim 10^{21}$~cm$^2$) of cold H$_2$ ($T \sim 100$~K),
and also evidence for the presence of warm/hot excited H$_2$ ($T \geq
500$~K). The H$_2$, observed by \fuse, probably corresponds to the
remnant of the molecular cloud in which the stars were formed and the
observed H$_2$ FUV lines do not originate in a CS disk. \cite{klr08a}
modeled the excitation conditions of H$_2$ and the \fuse\ spectra of
HBes stars using the {\it Meudon PDR Code} and showed that the
observed gas was located in relatively diffuse regions ($10-3000$
cm$^{-3}$) at significant distances from the central stars ($0.03-1.5$
pc).

None of the four targets with spectra of sufficiently high S/N showed
any evidence for H$_2$ emission. From the 3$\sigma$ upper limits to
the emission line flux, we calculated upper limits on the column
density and mass of H$_2$ for each star. We found that the column
densities should be lower than $\sim$10$^{23}$ cm$^{-2}$ at 150~K, and
lower than $\sim$10$^{21}$ cm$^{-2}$ at 1000~K, which does not
contradict the column densities derived from the \fuse\
observations. Upper limits to the masses of warm gas have been
estimated to be in the range from $\sim$10$^{-2}$ to $\sim$6 $M_{\rm
  Jup}$ (1 $M_{\rm Jup}$ $\sim$ 10$^{-3}$ $M_{\odot}$), assuming LTE
excitation, and depending on the adopted temperature.

It should be pointed out that mid-IR H$_2$ lines only probe warm gas
located in the surface layers of the observed media because of the
opacity of the interior layers when dust is present. The surface
layers of any potential disks surrounding our target stars do not
therefore contain sufficient warm gas to enable it to be detected in
emission at mid-IR wavelenghs. Following the calculations of
\cite{Carmona08}, the H$_2$ mid-IR lines produced by a gas-rich disk
with $T_{gas} = T_{dust}$ should not be observable with existing
instruments. As shown by \cite{Bitner07} and \cite{klr07a}, the warm
H$_2$ can be detected in the CS disk of a Herbig star, but to explain
the detection, particular conditions have to be assumed for the gas
and dust, such as $T_{gas} > T_{dust}$, which may be created by gas
heating by X-rays or UV photons. Our non-detections therefore imply
either that the gas and dust in any potential disks are well mixed and
have almost equal temperatures, or that these disks are simply not
present at all.


\begin{acknowledgements}
  This work is based on observations obtained at ESO/VLT (Paranal)
  with \visir, program number 076.C-0299. CMZ warmly thank M. Deleuil
  and J-C. Bouret for their help in preparing the observations. We
  thank X. Delfosse (LAOG) and the anonymous referee for their
  fruitful comments about the noise estimates. CMZ was supported by a
  CNES fellowship.
\end{acknowledgements}


\bibliographystyle{aa}
\bibliography{Martin-Zaidi_fin}


\end{document}